\documentclass[a4paper]{article}

\usepackage[english]{babel}
\usepackage[utf8x]{inputenc}
\usepackage[T1]{fontenc}

\usepackage[a4paper,top=3cm,bottom=2cm,left=3cm,right=3cm,marginparwidth=1.75cm]{geometry}

\usepackage{amsmath, dsfont, nicefrac}
\usepackage{graphicx}
\usepackage[colorinlistoftodos]{todonotes}
\usepackage[colorlinks=true, allcolors=blue]{hyperref}
\usepackage{natbib}

\title{Goodness-of-Fit Tests for Large Datasets} 
\author{Taras Lazariv and Christoph Lehmann \\ \small{ Center of Information Services and High Performance Computing (ZIH)} \\  \small{Technical University Dresden} }

\newcommand{\RR}{\mathds{R}}                

\newcommand{\eqdef}{\stackrel{\mathrm{def}}{=}}     
\newcommand{\iid}{\stackrel{\mathrm{i.i.d.}}{\sim}}  
\newcommand{\Ind}{\mathds{1}}               

\newcommand{\NIX}[1]{}                      

\NIX{
  Hinweise:
  \begin{itemize}
    \item Fuer Intervalle und Operatoren sollten eckige Klammern verwendet werden.
    \item Der Schaetzer fuer die Varianz wird mit $S^{2}$ bezeichnet.
    \item Der Schaetzer fuer die korrigierte Varianz wird mit $S^{*2}$ bezeichnet.
  \end{itemize}
}

\begin{document}
\maketitle

{\let\thefootnote\relax\footnote{{This work was supported by the German Federal Ministry of Education and Research (BMBF, 01IS14014A-D) by funding the competence center for Big Data ``ScaDS Dresden/Leipzig''.}}}

\begin{abstract}
  Nowadays, data analysis in the world of Big Data is connected typically to data mining, descriptive or exploratory statistics, e.~g.\ cluster analysis, classification or regression analysis. 
  Aside these techniques there is a huge area of methods from inferential statistics that are rarely considered in connection with Big Data. 
  Nevertheless, inferential methods are also of use for Big Data analysis, especially for quantifying uncertainty. 
    
  The article at hand will provide some insights to methodological and technical issues referring inferential methods in the Big Data area in order to bring together Big Data and inferential statistics, as it comes along with its difficulties. 
  We present an approach that allows testing goodness-of-fit without model assumptions and relying on the empirical distribution.
  Especially, the method is able to utilize information from large datasets.
  Thereby, the approach is based on a clear theoretical background. 
  We concentrate on the widely-used Kolmogorov-Smirnov test that is applied for testing goodness-of-fit in statistics. 
  Our approach can be parallelized easily, which makes it applicable to distributed datasets particularly on a compute cluster.
  
  By this contribution, we turn to an audience that is interested in the technical and methodological backgrounds while implementing especially inferential statistical methods with Big Data tools as e.~g.\ Spark. 
  
\end{abstract}

keywords: statistical inference, Big Data, Kolmogorov-Smirnov test, exploratory statistics, outlier detection

\section{Introduction}

Dealing with Big Data often is characterized by different aspects that are often referred to as the \glqq four Vs\grqq, namely volume, velocity, variety and veracity, cf.~\cite[p.\,373]{franke2016}. 
As it is difficult to define what is small, medium or big data, we follow the quite simple definition given by John Tukey for Big Data: \glqq anything that won’t fit on one device\grqq, cf.~\cite{donoho2017}.

Besides the pure amount of data, one need to decide if the available data are really suitable for the provided problem statement. Unfortunately this question cannot be answered by statistical means.
Another important concern is a selection bias, which means that the data are biased according to some effects, factors, unobserved constraints etc. 
A prominent example is the Google Flu Trend that failed to predict the A/H1N1 pandemic in 2009, caused by the selection bias. 
Not only this issue, but also some other interesting aspects of analyzing Big Data are discussed in~\cite{harford2014}.

Often Big Data is recorded casually and/or arises by combining different data sources, which is the aspect of veracity, cf.~\cite[p.\,374]{franke2016}. 
In order to use such data with inferential statistics, one should keep in mind that statistical methods are developed typically for planned experiments. 
This is strongly connected to a method's assumptions and the question of effects to investigate and how to measure them.
Thus, especially the methods from nonparametric statistics are of great interest because they have only weak assumptions.

The article at hand mainly covers the volume aspect in connection with inferential statistics.
Thereby, one aim is to bring together data mining and statistics, as both can benefit from each other. 

We want to briefly illustrate statistical theory for the Kolmogorov-Smirnov (KS) test for providing some insights when it might be beneficial to come up with inferential methods for large datasets. The KS test may be used to find anomalies or identify structures in the data.  

\bigskip


In \cite{sobolewski2013} some statistical tests are used to analyze data streams, in order to identify pattern changes (so called virtual concept drift).
In contrast to \cite{sobolewski2013}, we concentrate on the KS test only, but we also investigate methodological and implementational details while using a software tool like Apache Spark. 
Thereby, we discuss the application of the methods of inferential statistics, esp. statistical tests, to large volumes of data.
One of the first contributions that mentioned the issue of large datasets in connection with inferential statistics was~\cite{granger1998}, who claimed that a large amount of data might not be advantageous for inferential statistics. 
Especially for testing goodness-of-fit the problem of too high discriminatory power arises. As a consequence, the application of a statistical test does not make sense and the information from the data cannot be fully utilized. 

In this article we present an approach to use the KS test for a large amount of data, while not having the above mentioned problem.
Additionally, there is no model assumption needed, as we rely on the empirical distribution and therewith making use of a large dataset without the drawback of too high discriminatory power. 
For analyzing Big Data, typically there is no appropriate known distribution model that would fit the data, as there is no information about the \glqq true\grqq\ underlying process where the data comes from.
Our approach is general in the sense that it can be used with the most prominent goodness-of-fit tests, such as KS test and $\chi^2$ test. 
In order to keep it short, we will demonstrate it based on the KS test. 
We will concentrate on existing methods that are already implemented to some extent. 
That means, we focus on the perspective of statistical application using tools that are available and ready-to-use.
Basically, we suggest a transformation approach to turn a two sample problem into a one sample problem. 
By using a statistical test, it is possible to quantify the reliability/uncertainty of the method.
From the technical point of view, an implementation of the presented approach can be realized quite easily even for distributed data on a compute cluster. 

Overall, we address a twofold audience. On one hand the statistician who has to cope with large amounts of data that cannot be handled without stronger computational resources.
On the other hand, for interested readers from the field of computer science we want to provide some insight on the KS test as it might be of interest for more data-driven approaches having only weak assumptions on the data.


The article is structured as follows. 
After this introduction, in Section~\ref{sec:BD_challenges_solutions}, we sketch some important issues of using Big Data in connection with statistical analysis. 
Section~\ref{sec:theoretical-bg} deals with the theoretical background of the KS test and presents a transformation approach that reduces some arbitrary goodness-of-fit problem to testing for uniformity. 
This is beneficial as, e.~g.\ Apache Spark\footnote{\url{https://spark.apache.org}}, does not provide such a great amount of available distributions out of the box for the KS test as R\footnote{\url{https://www.r-project.org}} or Python\footnote{\url{https://www.python.org}} do. 
Section~\ref{sec:Application_Implementation} contains some simulations based on the methodology in Section~\ref{sec:theoretical-bg}. 
Furthermore, we discuss some issues that arise while using a large amount of data for statistical inference, esp. testing. 
Finally, the article ends with an outlook and conclusions in Section~\ref{sec:conclusions}.

\section{Big Data: Challenges and Solutions} \label{sec:BD_challenges_solutions}

Big Data comes along with many issues that affect the analysis workflow on the technical and methodological level as well. 
In the following we shortly discuss some of the most important ones and the implications.

One of the issues in connecting inferential statistics and Big Data is the assumption \glqq $N = \text{All}$\grqq, with $N$ denoting the sample size, which is discussed controversially.
It means that \glqq sampling does not make sense if all data is available\grqq, cf.~\cite[p.\,354]{vanAalst2016}.
On the other hand there is the position that \glqq $N = \text{All}$\grqq\ mostly is not true in important cases, cf.~\cite[p.\,19]{harford2014}.
As a consequence, \glqq $N = \text{All}$\grqq\ should be seen at least as an assumption that needs to be checked carefully.
In the case where one can assume that having all data means to have the full knowledge regarding the question of interest, the application of an inferential method is not of use at all, cf.~\cite[Section 2]{hand1999}. 
But also this case, especially for massive datasets, has its specialties that need to be considered, cf.~\cite[Section 12.1.1]{vanAalst2016}. 
In our article, we will not refer to this situation as sampling is not of use there. 

Nevertheless, working with samples and therewith, using inferential statistics, might be helpful in different situations.
One simple reason for sampling is a complexity reduction of data at least in terms of the data size. 
Note that this strategy will not work in the case of high dimensional data, where the number of variables exceeds the number of observations for every variable.
An interesting overview on the issue of high dimensional data is given in~\cite{fan2014}.

Data analysis also comes with the aspect of efficiency, i.~e.\ \glqq assuming we have infinite computing resources for Big Data analytics is a thoroughly impracticable plan, the input and output ratio (e.g., return on investment) will need to be taken into account [...]\grqq, cf.~\cite[p.\,23]{tsai2015}.
Another aspect is working with streams directly or samples out of streams, cf.~\cite[p.\,112]{leskovec2014}. 
Sampling in streams is not an easy task as the population size grows continuously and typically one will not store all the streamed data, cf.~\cite{park2007, bhattacharya2017}. 
Note that the sampling issue typically has implications on the often presumed iid-assumption (independent and identically distributed random variables).
Further aspects on working with samples can be found in~\cite[Section 4.2]{srivastava2015}.

Nevertheless, given that the assumptions for an inferential method are met, one might interpret results of a statistical test on these data in a descriptive/exploratory manner, i.~e.\ the results only refer to the analyzed dataset.
From a practical point of view, a statistical test can provide a descriptive or exploratory indication for some issue that we are interested in. 
This reflects the more pragmatic view from the area of data mining, as one is looking for tools and methods that are of practical use, cf.~\cite[Section 2]{hand1999}.
Unfortunately, many of the assumptions for inferential methods are too strong, i.~e.\ these are not met while working with Big Data, e.~g.\ an assumption for normal distribution of the data. 
Thus, the area of nonparametric (or distribution free) statistics is of great interest here, because only some mild assumptions need to be fulfilled.
The KS test belongs to the class of nonparametric tests.



\bigskip

Beside methodological issues, the application of statistical methods for large data sets comes along with the technical issue of parallelization as this forms a base of getting advantage by using a compute cluster. 
Basically, parallelization means that a method, e.~g.\ a statistical test, can be broken into single tasks that are performed independently and simultaneously on different CPUs or nodes in a compute cluster and on subsets of data.
Note that this might be a challenging problem as it depends not only on the method itself, but also on programming languages, hardware architectures etc. 
Insights on the methodological side for parallelization are provided in \cite{jordan2013}. 
The technical aspects are discussed in the vast literature about high performance computing and parallel programming, for an introduction cf.~\cite{pacheco2011}.

\section{Theoretical Background and Statistical Methodology}\label{sec:theoretical-bg}

In this section, the most important statistical theory regarding the one-sample KS test is presented. 
Often, there is a need to compare two samples whether they could stem from the same distribution or not. 
In connection with testing, this is a two sample problem.


\subsection{Kolmogorov-Smirnov Test in Brief}

In the following, we provide a short overview on the KS test. A more general discussion on goodness-of-fit tests can be found e.~g.\ in \cite[pp.\,335]{lehmann2004}.
For more detailed information on the KS test cf.~\cite[pp.\,584]{lehmann2005} or \cite[pp.\,108]{gibbons2011}. 

Let $X_1, \ldots, X_n$ be a random sample of i.i.d.\ real-valued observations from an unknown, continuous distribution with the cumulative distribution function (cdf)~$F_X$. 
The KS test is a goodness-of-fit test to investigate the hypothesis
\begin{equation}
  H_0: F_X = F_0 \quad	\text{vs.} \quad H_1: F_X \neq F_0,
\end{equation}
where~$F_0$ denotes the cdf of a fully specified theoretical distribution. 
Roughly speaking, the test checks whether a random sample of data can be assumed to stem from the specified theoretical distribution~$F_0$. 
The test statistic $T_n$, the KS statistic, is defined as
\begin{equation}\label{eq:KS-statistic}
  T_n \eqdef \sqrt{n} \, \sup\limits_{x \in \RR}  |\hat{F}_n(x) - F_0(x)|,
\end{equation}
where~$\hat{F}_n$ denotes the random empirical cumulative distribution function (ecdf)
\[
  \hat{F}_n (x) \eqdef n^{-1} \sum\limits_{i=1}^n \Ind_{]-\infty, x]} (X_i), \quad x \in \RR.
\]
As $n \to \infty$, the limiting distribution of $T_n$ is the so called Kolmogorov distribution with distribution function
\[
  P(T_n \leq t) = 
  \begin{cases}
    0, & t \leq 0 \\
    1 - 2 \sum\limits_{j=1}^\infty (-1)^{(j-1)} \exp\left( -2 j^2 t^2\right), & t > 0.
  \end{cases}
\]
This asymptotic test rejects the null hypothesis if $T_n > k_{1-\alpha}$, with $k_{1- \alpha}$ denoting the $1-\alpha$ quantile of the Kolmogorov distribution. 
The significance level is set by selecting an appropriate value for the probability $\alpha \in \, ]0,1[$.
Having large sample sizes, the usage of the asymptotic version of the KS test is justified.
The KS test is consistent against the alternatives (which is a desirable property of a test), i.~e.\ the probability for rejecting a wrong null hypothesis converges to one as $n \to \infty$ (cf.~\cite[p.\,342]{lehmann2004}). 
In other words, the discriminatory power between two different distributions of the KS test increases with increasing sample size. 
Note that calculating the power against alternatives varies depending on the investigated alternatives, cf.~\cite{boyerinas2016}, \cite{senger2013}, \cite{razali2011} or \cite{milbrodt1990}. In that sense, the behavior of testing power is strongly connected to the concrete application. 

\subsection{Transformation to Uniformity}

Here we transform the two sample goodness-of-fit problem into testing uniformity as a one-sample test.

For the first step, assume two independent random samples $X_i \iid F_X$, $i = 1, \ldots, n$, and $Y_j \iid G_Y$, $j = 1, \ldots, m$, whereby $F_X$ and $G_Y$ denote some arbitrary unknown, but continuous and real valued cdfs. 
Now, we want to check whether our random samples could stem from the same distribution, i.~e.\ we could test the hypothesis
\begin{equation}\label{eq:hypothesis-two-sample}
  H_0: F_X = G_Y \quad	\text{vs.} \quad H_1: F_X \neq G_Y.
\end{equation}
This two sample KS test can be found e.~g.\ in~\cite[pp.\,234]{gibbons2011}.

From probability theory we know the so called probability integral transformation, cf.~\cite[p.\,54]{casella2002}. 
Let a random variable~$Z$ have a continuous cdf denoted by~$F_Z$. 
Then, the transformation~$F_Z(Z)$ has a uniform distribution~$U(0,1)$ on the unit interval~$[0,1]$, denoted as 
\begin{equation}\label{eq:transformation-uniform}
  F_Z(Z) \sim U(0,1).
\end{equation}

The random ecdfs 
\begin{align*}
  \hat{F}_n (x) \eqdef n^{-1} \sum\limits_{i=1}^n \Ind_{]-\infty, x]} (X_i) \quad \text{and} \quad \hat{G}_m (x) \eqdef m^{-1} \sum\limits_{j=1}^m \Ind_{]-\infty, x]} (Y_j), \quad x \in \RR
\end{align*}
are consistent estimators for the theoretical cdfs $F_X$ and $G_Y$ resp. in \eqref{eq:hypothesis-two-sample}.
Based on \eqref{eq:transformation-uniform} it holds
\begin{align}
  \hat{F}_n (X_i) \iid U(0,1), \ \forall i \quad \text{and} \quad \hat{G}_m (Y_j) \iid U(0,1), \ \forall j.
\end{align}
Note that independence of the transformed samples is still maintained, which follows from the disjoint blocks theorem, cf.~\cite[p.\,76]{karr1993}.
The two sample problem \eqref{eq:hypothesis-two-sample} can be turned into a one-sample problem as under $H_0$
\begin{equation}\label{eq:convergence-to-uniformity-under-H0}
  \hat{F}_n (Y_j) \stackrel{d}{\longrightarrow} U(0,1) \ \text{and} \ \hat{G}_m (X_i) \stackrel{d}{\longrightarrow} U(0,1)
\end{equation}
as $n, m \to \infty$, where $\stackrel{d}{\longrightarrow}$ stands for convergence in distribution. 

Practically speaking, for observed data, i.~e.\ two sample realizations $x_1, \ldots, x_n$ and $y_1, \ldots, y_m$ we take one of the realized ecdf, say for the estimator $\hat{F}_n$, and calculate the transformation
\begin{equation}\label{eq:transformation-data-into-one-sample}
  n^{-1} \sum\limits_{i=1}^n \Ind_{]-\infty, y]} (x_i), \quad y \in \{y_1, \ldots, y_m\}.
\end{equation}
Under $H_0$ from \eqref{eq:hypothesis-two-sample}, we would expect the transformed data from \eqref{eq:transformation-data-into-one-sample} to be approximately distributed as $U(0,1)$, which can be checked by using the one-sample KS test as already presented.
Note that the transformation from \eqref{eq:transformation-data-into-one-sample} does not need any sorted data, which makes it easy to implement this step even for distributed data on a compute cluster. 
Remark: The two sample version of the KS test typically needs sorted data (sorting data is computationally intensive) for getting the supremal distance between the two ecdfs, cf.~\cite[pp.\,617]{press1992}. 

\bigskip
In Section \ref{sec:theoretical-bg}, we showed that it is sufficient to concentrate on testing uniformity as a goodness-of-fit test problem can be transformed adequately. 
As such a transformation strategy is common in statistics, alternatives for testing uniformity were investigated broadly in the literature using different approaches. 
An overview of different tests with an investigation of some of their properties can be found e.~g.\ in \cite{blinov2014}, \cite{marhuenda2005}, \cite{gokhale1983} or \cite{dudewicz1981}.
Here we have chosen the KS test as it is one of the most popular tests for goodness-of-fit, especially in the case of continuous distributions.

\section{Application and Implementation}\label{sec:Application_Implementation}

In our application we are interested in detecting anomalies by using the KS test. Basically, this means to compare two distributions in order to investigate their equivalence.
In this section the simulation results are described in a first step. 
In a second step the results are discussed in a more general manner, esp.\ referring the application of statistical tests with large datasets.

\subsection{Simulation Results}\label{sec:Application_Simulation}

For illustrating the above presented transformation approach, we performed simulations of the KS test. 
Thereby, we check the hypothesis whether two samples could stem from the same distribution, according to \eqref{eq:hypothesis-two-sample}.
Having in mind the idea of anomaly detection, there might be information about a common state for some system that we will name reference distribution with cdf $F_X$. 
On the other hand, there is another distribution with cdf $G_Y$ that might characterize the above system over a time period, e.~g.\ during the last hour or the last day. This distribution we will name comparison distribution.
Now we are interested, whether there is some indication that the system does not behave as usual, i.~e.\ $F_X \neq G_Y$.

\begin{figure}[!tbp]
	\centering
	\begin{minipage}[b]{0.49\textwidth}
		\includegraphics[width=\textwidth]{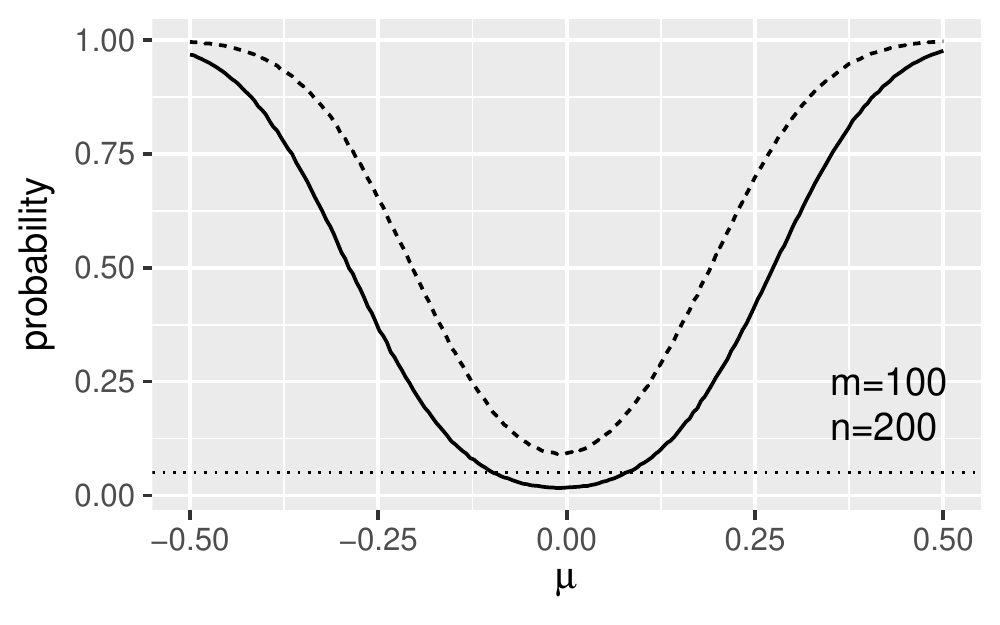}
	\end{minipage}
	\hfill
	\begin{minipage}[b]{0.49\textwidth}
		\includegraphics[width=\textwidth]{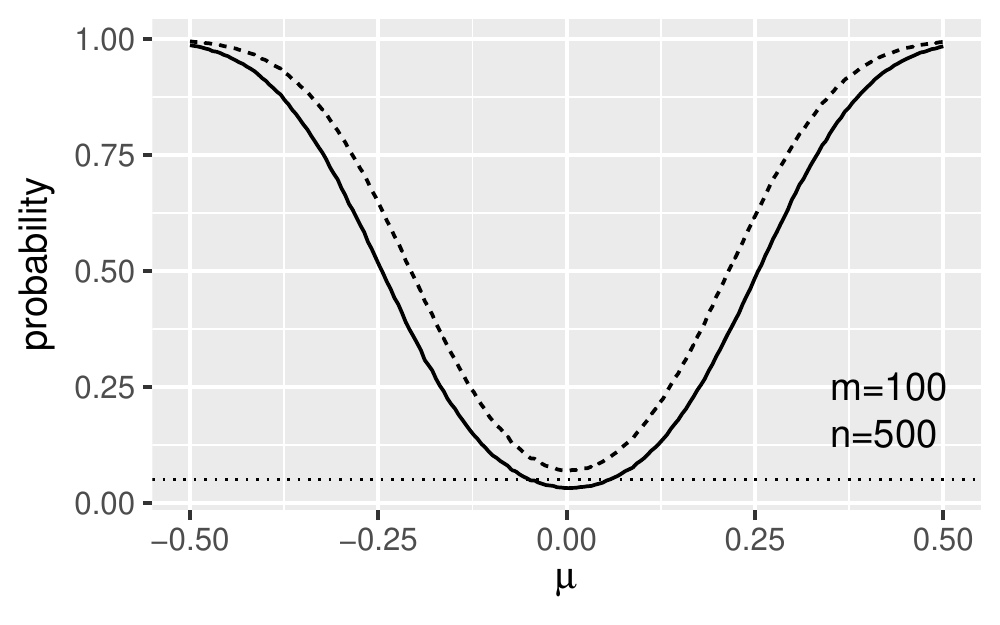}
	\end{minipage}
	\begin{minipage}[b]{0.49\textwidth}
		\includegraphics[width=\textwidth]{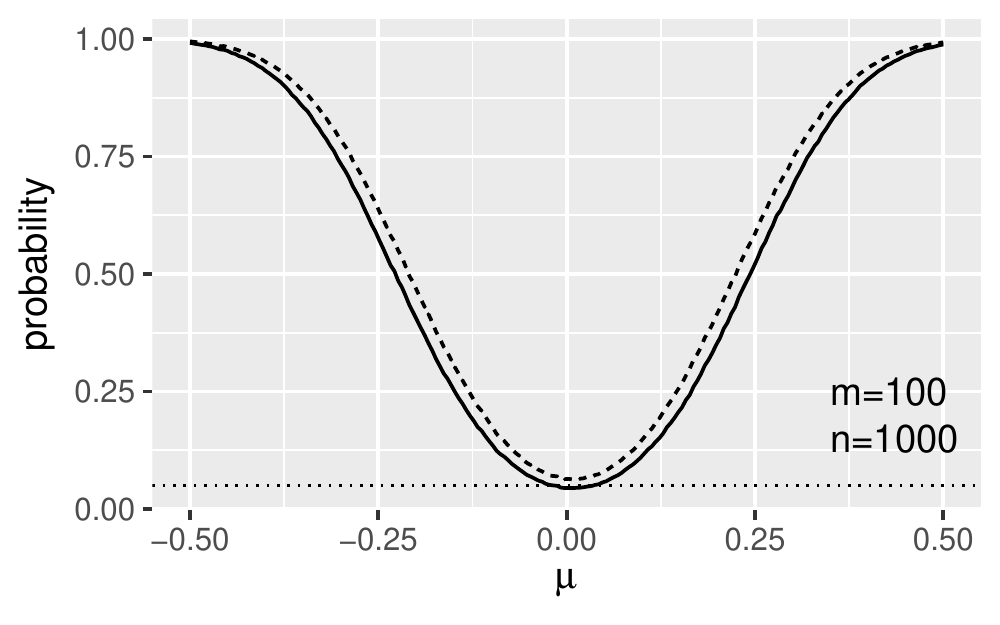}
	\end{minipage}
	\hfill
	\begin{minipage}[b]{0.49\textwidth}
		\includegraphics[width=\textwidth]{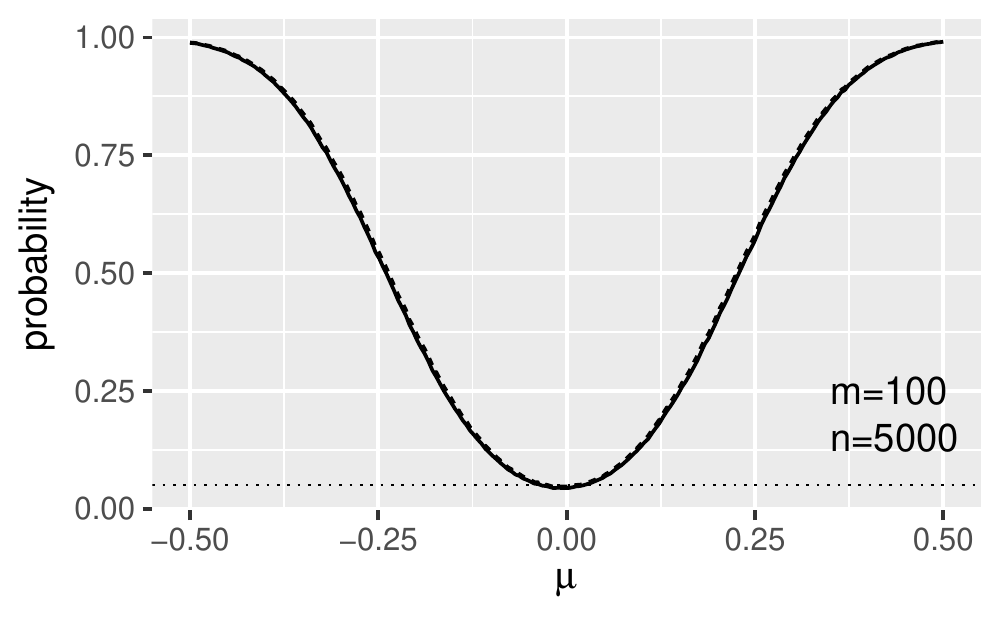}
	\end{minipage}
	\caption{The influence of the reference sample size, with a fixed comparison sample size, onto the empirical power function of the two sample KS-test (dashed line) and transformation approach (solid line) for a standard normal distribution. The dotted line represents the significance level $\alpha=0.05$.}\label{fig:PowerFunction_normal_ratio}
\end{figure}

\begin{figure}[!tbp]
	\centering
	\begin{minipage}[b]{0.49\textwidth}
		\includegraphics[width=\textwidth]{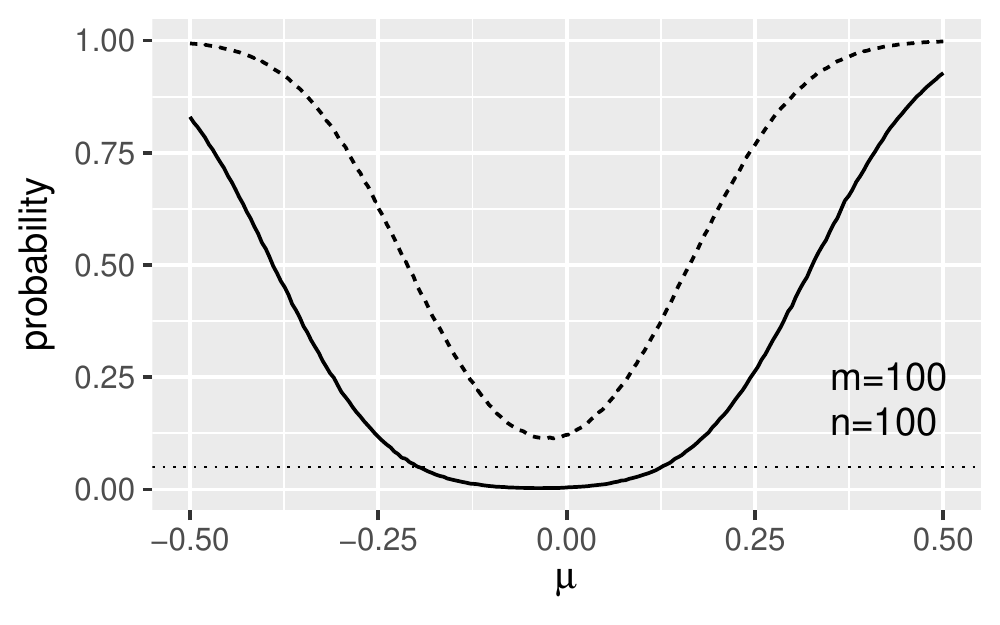}
	\end{minipage}
	\hfill
	\begin{minipage}[b]{0.49\textwidth}
		\includegraphics[width=\textwidth]{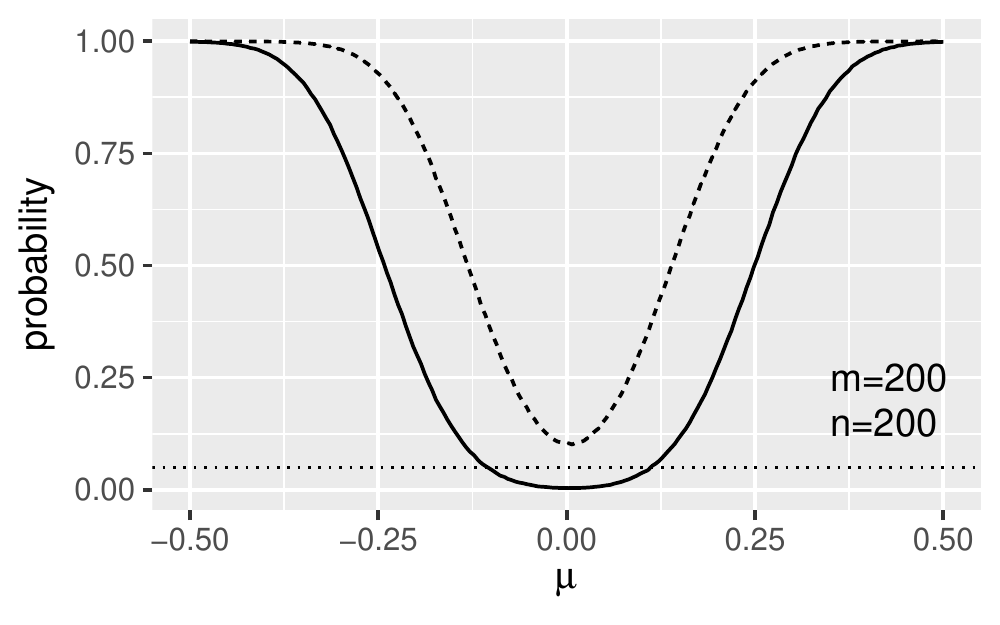}
	\end{minipage}
	\begin{minipage}[b]{0.49\textwidth}
		\includegraphics[width=\textwidth]{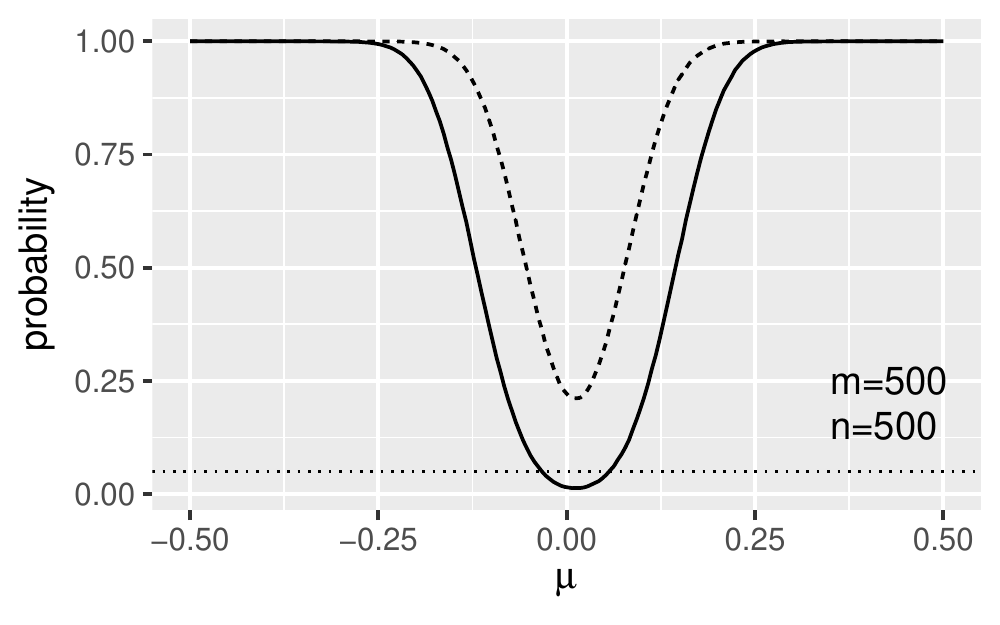}
	\end{minipage}
	\hfill
	\begin{minipage}[b]{0.49\textwidth}
		\includegraphics[width=\textwidth]{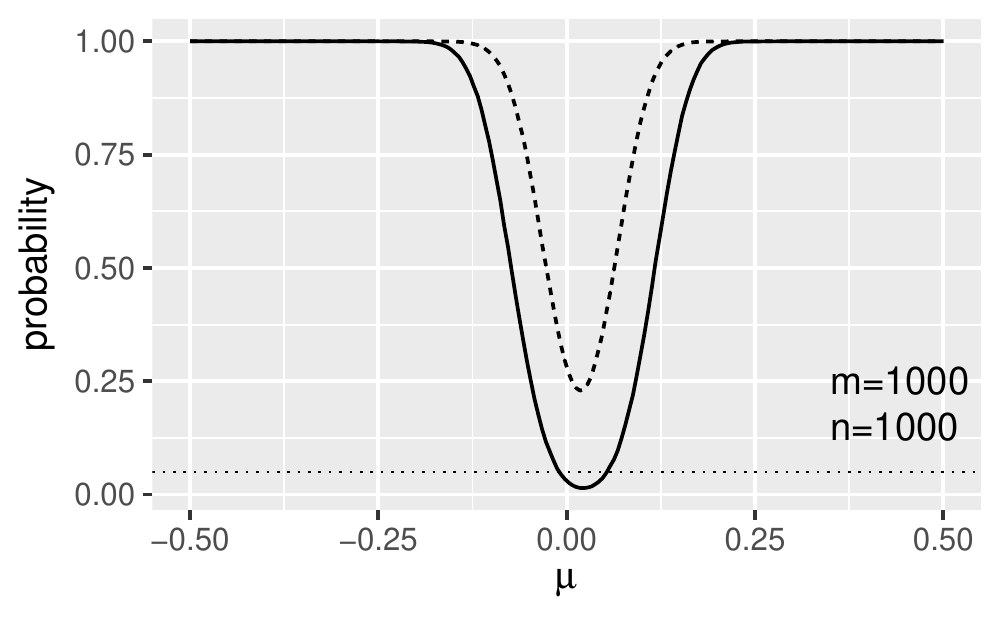}
	\end{minipage}
	\caption{The influence of sample sizes with a fixed ratio for comparison and reference sample onto the empirical power function of the two sample KS-test (dashed line) and transformation approach (solid line) for a standard normal distribution. The dotted line represents the significance level $\alpha=0.05$.}\label{fig:PowerFunction_normal_size}
\end{figure}

Figures~\ref{fig:PowerFunction_normal_ratio} and~\ref{fig:PowerFunction_normal_size} illustrate the empirical power functions of our simulations for varying values of the sample sizes $m$ and $n$, where the transformation approach is compared to the two sample KS test. 
Basically, the power function of a statistical test provides the probability to reject the null hypothesis (no matter whether it is correct or not).
Here we have chosen a standard normal distribution, $N(0,1)$, under the null hypothesis, characterized with cdf~$F_X$ (according to \eqref{eq:hypothesis-two-sample}). 
Furthermore, the space of the alternative hypothesis~$H_1$ contains all normal distributions $N(\mu, 1)$ with $\mu \neq 0$.
Thus, the power functions depend on the parameter~$\mu$ of a normal distribution~$N(\mu, 1)$.
Iterating over fixed values of $\mu \in \RR$, each statistical test (here: two sample KS test and transformation approach) is performed $10^5$ times. 
Depending on $\mu$, the empirical power function is the relative frequency of cases where $H_0$ was rejected during the $10^5$ replications. 


The illustrated power functions only reflect probabilities referring alternative hypotheses that contain the normal distribution. 
In that sense, esp.\ the type~II error (accept a wrong null hypothesis) only reflects deviations from the parameter~$\mu$ of the normal distribution, but not deviations with respect to some arbitrary distribution. 
More precisely, the illustrated empirical power functions are estimates for these probabilities based on the relative frequencies of rejecting the null hypothesis.
As it is desirable, the rejection probability is increasing while the deviation from $H_0$ increases.
As expected, the power functions are steepening with increasing sample sizes, this is 
the consistency of the KS test (as mentioned in Section~\ref{sec:theoretical-bg}).
The chosen level of significance reflects the reliability referring the decision, i.~e.\ it is a possibility to handle the uncertainty that might arise from the data, namely the sample error.
Another simulation setting was performed with an exponential distribution under $H_0$, since it reflects a non-symmetrical case. As this led to quite similar results, these are not reported here.

Note that transformations based on the reference distribution, i.~e.\ its ecdf, lead to testing of discrete distributions. 
The KS test becomes conservative as it is applied to discrete distributions, i.~e.\ the level of significance becomes no larger than in the case of a continuous distribution, cf.~\cite{goodman1954}.
Roughly speaking it is more unlikely to reject the null hypothesis. 
This can be seen in Figures~\ref{fig:PowerFunction_normal_ratio} and~\ref{fig:PowerFunction_normal_size} as the power function of the transformation approach is shifted downwards referring the two sample test.
While the type~I error (rejecting a true null hypothesis) is decreasing, the type~II error is increasing.
The type~I error is directly accessible via the (empirical) power function, the type~II error is calculated as one minus the power function.
This shift is vanishing as the sample size for the reference distribution grows.
Another aspect that comes up by using an ecdf for transformation are ties, because an ecdf is a step-function consisting of constant parts and therewith creating ties within the transformation.
To get rid of occurring ties, one could add some random noise to the transformed values, whereby one need to ensure that the transformed values are not exceeding the unit interval.

Note the twofold approximation in connection with the transformation idea. 
As we are using a one-sample test, the distribution $F_0$ is estimated by $\hat{F}_n$ but the test \glqq sees\grqq\ $\hat{F}_n$ as the \glqq true\grqq\ distribution. 
For getting the properties of a one-sample KS test we consider $n \to \infty$, while the size $m$ of our sample to be compared remains unchanged or $m$ is increasing more slowly at least.
If this relation is violated, nearly every $H_0$ will be rejected. 
The bottom line is the transformation approach works in the case of a sufficient large sample size $n$ from the reference distribution and the sample size $m$ of the distribution to be compared should not be too large.
This effect is illustrated in Figure~\ref{fig:PowerFunction_normal_size}. 
For practical applications mostly it is possible to get increasing information about the reference distribution while we have only a smaller amount of information about the distribution(s) to be compared.  

As a consequence, the ratio of sample sizes $r \eqdef \nicefrac{m}{n}$ is an important aspect.
Based on the above simulation results, see Figure~\ref{fig:PowerFunction_normal_ratio}, we recommend to strive for $r < 0.2$. 
Thereby, one might be aware that a small sample size~$m$ of the comparison sample may be not in contradiction to many distributions.
This is closely connected to the fact, that the KS test with the transformation approach depends on~$m$ and not on the sample size~$n$ of the reference sample.
The critical value of the KS test depends on $m$ while the sample size~$n$ only influences the precision of the estimated reference distribution. 
For practical use, one should have a sufficient large sample for the reference distribution that is able to characterize its important characteristics.
In a second step, based on the above recommendation ($r < 0.2$), a maximum sample size~$m$ can be chosen.
Note that based on the transformation into a one sample problem, the reference distribution is emphasized against the comparison distribution. 
This is in contrast to the two sample test, where the information of both samples are weighted equally in some sense.

\subsection{General Discussion on Statistical Tests with Big Data}\label{sec:statistical-tests-and-BD}

In statistical inference mainly the sample error is involved in nearly every consideration. 
Typically, a sufficient large amount of observations is desirable in order to get more reliable results.
As a consequence of having Big Data in connection with goodness-of-fit, already small differences between two datasets are indicated as statistically significant, i.~e.\ the null hypothesis is rejected. 
This aspect refers to the discriminatory power of test, which increases with an increasing sample size as already mentioned in Section~\ref{sec:theoretical-bg} and the simulations in Section~\ref{sec:Application_Simulation}.
Unfortunately, such a result does not reflect the practical relevance of 
an existing difference/distance. 
Up to the present, this issue of interpreting test results often is not reflected sufficiently in practice, cf.~\cite{gupta2012} or \cite{ranganathan2015}.
Especially in connection with Big Data (but not only there), we need to define a difference/distance between two distributions in advance that might still be acceptable from the practical point of view. 
The statistical test then evaluates whether a statistically significant deviation from this defined distance exists.
Thereby, the application of confidence bands for distribution functions might be useful in order to ease the single KS test runs.
These confidence bands can be constructed based on the KS test statistic and its asymptotic distribution, cf.~\cite[pp.\,121]{gibbons2011}.

In many practical applications there might be no appropriate probability distribution model because of \glqq unusual\grqq\ distribution shapes that make it hard to apply some known distribution model. 
Thus, a utilization of the ecdf directly is justified (at least as reference distribution), because having a large amount of observations makes this ecdf a reliable reference of an unknown data generating process. 
Thereby, we need to carefully choose our dataset in such a way that it includes what we are interested in. 
Especially referring Big Data, this is an issue as the datasets are not arising from a clear sample design, but are often observational data (for this issue cf.~\cite[Section 2.3, p.\,374]{franke2016}).
As a consequence, at best there is little control about the data generating process.

\subsection{Technical Details}

From the technical point of view we need to distinguish two cases. 
The first is applying a test to one long sample, which is typically a job for Big Data frameworks as the data do not fit into memory of a single machine.
The second case is applying a test to many samples simultaneously, whereby every single sample fits into memory.
This problem can be solved either with the help of parallel programming techniques or with Big Data frameworks.

Testing only smaller amounts of data does not need the possibility of distributed data and therefore utilizing a tool like Spark. 
Nevertheless, performing a huge amount of test runs in parallel, Spark can be used but running every single test on one compute node. 
Thereby, the Scala implementation of the KS test can be used. 
Remark: The RDD-based implementation for Spark uses a part of the Scala implementation of the KS test, esp.\ for obtaining the p-value from the calculated realization of the test statistic. 
With a fixed dataset we performed a rash of KS tests on Spark and R as well, which led to quite similar results (up to seven decimal places). 
Moreover, there is a difference regarding the implementation for the KS test. 
While an implementation of the one sample test is quite easy, the two sample version typically needs sorting of the data that might be computationally intensive esp. with large datasets. 
From this point of view, it is of advantage to utilize an approximation by transforming the problem.
Additionally, the transformation as suggested is easy to implement even in the case of distributed data as it can be parallelized. 

Typical software used for data analysis, like R, Python, SPSS, etc., is not capable of dealing with large datasets out of the box. 
The reason is, that by design, these tools are keeping all the data in the main memory of the computer system, which is very limited in size. 
There are solutions available, such as, e.g.\ the R package \verb|bigmemory|, that does not read the whole data into the memory, but rather references some part of it in cache and loads other portions of it on demand. 
Python can handle larger datasets using packages like \verb|pandas| (working with chunks) or \verb|dask| (distributed datasets). 
Even if this problem is solved, another one arises, namely one-threaded computations even on multicore systems.  
A solution would be to parallelize the execution, using different extensions, like \verb|parallel| in R, or packages like \verb|multithreading| or \verb|multiprocessing| in Python. 
If none of the above helps, one should consider switching to the Big Data Frameworks. 
The most widely used ones come from Apache Software family, e.g. Apache Hadoop\footnote{\url{https://hadoop.apache.org}}, Apache Spark, etc.
The main advantage of those are distributed computing and data storage, fault tolerance and easy scalability.

The above mentioned software tools are developed by different communities, that have their special core aspects. 
E.~g.\ R is mainly used and developed within a statistical motivated community. 
Thus, R has a wide coverage of implemented statistical methods, especially in inferential statistics. 
Python is used often in the engineering field and has well developed libraries especially for tasks in deep learning. 
Upcoming frameworks as Apache Spark or Apache Flink\footnote{\url{https://flink.apache.org}} are developed by a community that comes from the field of computer science. 
Compared to a tool like R, these frameworks are quite new developments that focus on the problem to work with large amounts of data. 
Thereby, these frameworks are mainly focusing on basic descriptive and exploratory statistics but may lack of inferential methods. 
Apache Spark Vers.~2.3.2 comes with a KS test implementation that is restricted to the two-sided one-sample test with the uniform and the normal distribution as reference distributions under the null hypothesis. 
With the above presented transformation approach it is possible to use this one-sample variant of the KS test in connection with the uniform distribution. 

\section{Conclusions and Outlook}\label{sec:conclusions}

Often for Big Data there is no appropriate distribution model to represent the data.
Additionally, large sample sizes for testing are not desirable in some sense.
In order to avoid these problems our proposed transformation approach even utilizes information from a large dataset whereby no distributional assumption is required.
Based on the theoretical background we developed our method and performed simulations in order to demonstrate its properties. 
The method allows a more data-driven approach whereby it remains possible to assess its uncertainty that arises from a sample.
In order to apply our transformation approach, a one sample implementation of the KS test that allows for testing uniformity is sufficient. 

From the technical point of view the above approach can be implemented for distributed datasets which makes it applicable to Big Data, particularly on a compute cluster.

\bigskip
Furthermore, working with ecdfs and the transformation into uniformity testing constitutes a unified measure that may allow to compare deviations for different comparison samples over different reference distributions.
But this issue might be investigated in a further work. 
Testing uniformity has a prominent position due to its generality which leads to many extensions of our approach, as it is not restricted to KS and $\chi^2$ tests. 
Furthermore, there are possibilities to expand, e.g.\ the KS test, to the multivariate case, cf.~\cite{justel1997}, \cite{sugiura1965} or \cite{fasano1987}. The same question arises for $\chi^2$ test and testing uniformity.
On the technical field of investigation one could explore these alternatives referring their appropriateness for a parallelized implementation.

\newpage
\bibliographystyle{dcu}
\bibliography{reference}


\end{document}